\documentstyle{article}
\begin{document}

\input epsf

\title{Theory for Characterization of Hippocampal Long-Term
Potentiation Induced by Time-Structured Stimuli
}
\author{
\vspace{5.0mm}
M. Tatsuno\thanks{e-mail: masami@aizawa.phys.waseda.ac.jp} 
and Y. Aizawa \\
Department of Applied Physics, Waseda University \\
3-4-1, Okubo, Shinjuku, Tokyo 169, JAPAN}
\date{(Received November 29, 1996)}
\maketitle

\begin{abstract}
We theoretically investigate long-term potentiation (LTP)
in the hippocampus using 
a simple model of a neuron stimulated by three different
time-structured input signals (regular, Markov, and
chaotic).  The synaptic efficacy change is described
taking into account both N-methyl-D-aspartate (NMDA) and non-NMDA
receptors.  The experimental results are successfully
explained by our neuron model, and the remarkable fact that
the chaotic stimuli in the nonstationary regime produce the
largest LTP is discussed.
\end{abstract}

\section{Introduction}
One of the main problems in neuroscience is determining how
information is represented in the brain.  There have been many
experimental and theoretical studies in this direction in
the last few
years.  For example, Aertsen and Gerstein conducted a series of
experiments, and proposed the concept of ``effective
connectivity''~\cite{rf:1}.
Tsukada et al. showed experimentally that the structure
of the inter-spike-interval has a significant effect on the
amplitude of long-term potentiation (LTP)
in the hippocampus of guinea-pigs,~\cite{rf:2} and
Fujii et al., based on a large number of experimental data,
theoretically proposed the hypothesis of dynamical cell assembly
in the central nervous system~\cite{rf:3}.
However, how information is represented
in the brain has not yet been elucidated because the
brain is a very complicated and hierarchically organized
system.  It is necessary to acquire more data about information
representation at the neuron level as well as at the network level.
It should also be emphasized that determination of how memory
information is represented in the brain is essential for
understanding of higher
brain functions, because all higher brain functions are based on 
coding of memory information.
Furthermore, not only experimental studies but also
theoretical studies, involving use of simple biological
neuron models or neural networks,
are necessary for elucidation of the mechanism of memory
information representation and its algorithm.

The purpose of this work was to investigate the effect of
the time structure of spike trains on the LTP amplitude in
the hippocampus using a simple
biological neuron model, and explain the basic synaptic
mechanisms underlying memory information representation.
For this purpose, we
constructed a single neuron
model considering both N-methyl-D-aspartate (NMDA) and non-NMDA receptors.
Then we applied three different time-structured
stimuli (regular, Markov, and chaotic) to our neuron model,
and compared
the LTP amplitude among the three cases.  A
striking finding was that the 
chaotic stimuli in the nonstationary regime produce the
largest LTP.  This suggests
the significance of nonstationary chaotic information
representation in memory information processing.

\section{Model}
Neurons in area CA1 of the hippocampus are typical examples
of neurons which
exhibit LTP~\cite{rf:4,rf:5}.  Since their physiological
properties have been extensively studied experimentally, we can
construct a single 
neuron model of this area.  Here, we adopt a discrete time
model with a
continuous membrane potential, a setup, however, which compromises
the biological plausibility and the simulation time cost.
We take into account the channel properties in the synaptic
region.

The basic time evolution of the activity of a single neuron is given by
\begin{equation}
O(t+1)= \left \{ \begin{array}{ll}
1, & \mbox{with prob. $f(V(t))$}\\
0, & \mbox{with prob. $1-f(V(t))$}
\end{array}, \right.
\end{equation}
where $O(t+1)$ is the output of the neuron at time $t+1$, $V(t)$ is
the membrane potential at time $t$, and
$f(\cdot)$ is a probability function which describes the relation
between the
membrane potential and the output.  Here, we adopt the following
form for $f(\cdot)$:
\begin{equation}
f(V(t))=\frac{1}{1+e^{-(V(t)+|V_{th}|)}},
\end{equation}
where $V_{th}$ is the threshold membrane
potential ($V_{th} < 0$) and $f(V(t))=1/2$ 
at $V(t)=V_{th}$.  

The membrane potential $V(t)$ can be decomposed into three
parts, i.e.,
\begin{equation}
V(t)=I(t)+V_{r}+\delta(t),
\end{equation}
where $I(t)$ is the post-synaptic potential induced by the
input current
at time $t$, $V_{r}$ is the constant resting membrane
potential, and $\delta(t)$ is
a small fluctuation of the membrane potential at time $t$
where the average value of $\delta(t)$ is zero.
The synaptic input through non-NMDA receptors $I(t)$
can be represented as follows: $I(t)$ depends on both the
synaptic efficacy $E(t)$ and
the time course of the input stimulus $\sigma (t)$, which
can be written as
\begin{equation}
I(t)=\sum_{n=- \infty}^{t}E(n)\sigma(n)e^{-\alpha (t-n)},
\end{equation}
where $E(n)$ is the synaptic efficacy at time $n$,
$\sigma(n)$ is the spike train of the input stimuli
which represents the existence of the input spike at
time $n$ 
($\sigma(n) = 1$ with input, $\sigma(n) = 0$ without input), and
$\alpha$ is the decay constant for non-NMDA receptors.  The
time course of $\sigma(n)$ determines the statistical
features of the input signals, such as
the average frequency or the average interval of the input
stimuli.

Now, let us consider LTP of this model in terms 
of the change of $E(t)$.
Physiologically, both depolarization of
the membrane and
 calcium entry into the post-synaptic neuron are necessary
for LTP to occur~\cite{rf:4}.  Therefore, we introduce the
variable $S(t)$, which
represents the concentration of calcium ions in the post-synaptic
neuron and is expressed as:
\begin{equation}
S(t) = C\sum_{n=- \infty}^{t}\{V(n)+|V_{L}|\}\theta(V(n)+|V_{L}|)e^{-\beta (t-n)},
\end{equation}
where $C$ is a constant, $\theta(\cdot)$ is a step function,
$V_{L}$ is the membrane potential required for removal of magnesium
ions from NMDA receptors ($V_{L} < 0$), and $\beta$ is the decay constant
for NMDA receptors.

Although it is not well known how the calcium-dependent
second messenger system produces LTP, several
experimental observations have revealed strong correlation
between the LTP amplitude and the frequency of the input
stimuli, that is, it is physiologically
plausible that (1) regular stimuli at 3 - 10 Hz can induce
LTP,~\cite{rf:2} but
(2) low-frequency stimuli ($\leq$ 1Hz) induce
long-term depression (LTD)~\cite{rf:6}.  We also need to
consider that the value
of $E(t)$ saturates in the regime of large $S(t)$ and $E(t)$ 
due to the limited number of
neurotransmitters and receptors.  Therefore, we propose the
following model for the change of
the synaptic efficacy $\Delta E(t)$:
\begin{eqnarray}
\Delta E(t) & = & E(t+1) - E(t) \nonumber \\
& = & C_{1}(1-e^{-C_{2}(S(t)-C_{3})^{2}}) \nonumber \\
&{}& \times
\left(\frac{1}{1+e^{-C_{4}(-(E(t)-C_{5}))((S(t)-C_{6})-C_{7})}}\right)
\nonumber \\
&{}& \times
\left(\frac{1}{1+e^{-(S(t)-C_{3})}}-C_{8}e^{-C_{9}(S(t)-C_{10})^{2}}\right),\nonumber \\
\end{eqnarray}
where $C_{1}, C_{2}, C_{3}, C_{4}, C_{5}, C_{6}, C_{7},
C_{8}, C_{9}$, and $C_{10}$ are all parameters.  A schematic
 of $\Delta E(t)$ is given in Fig. 1, where the values 
of $C_{i}$
are fixed in our simulations.

\begin{figure}
\epsfxsize=15.9cm \epsfysize=11.0cm \epsfbox{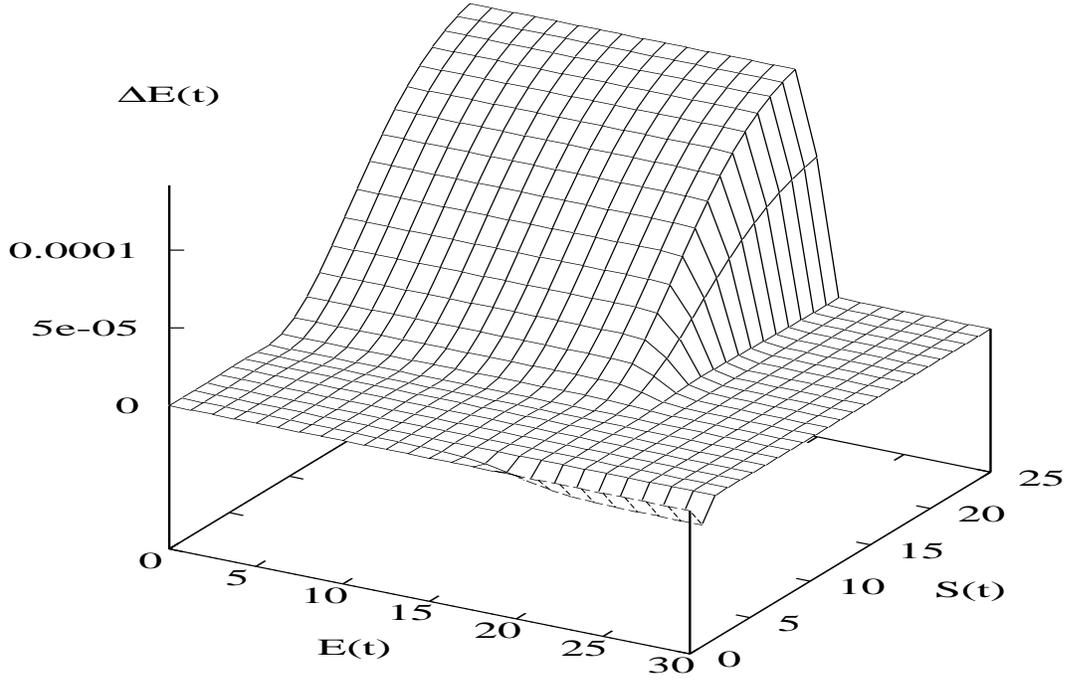}
\caption{$\Delta E(t)$ as a function of $E(t)$ and
$S(t)$.}
\end{figure}
For $E(t) \,\,\,\vphantom{(}\lower4pt\hbox{$\stackrel
{ \displaystyle {<}}{\sim}$}\,\,\, 20$ and $S(t) \,\,\,\vphantom{(}\lower4pt\hbox{$\stackrel
{ \displaystyle {>}}{\sim}$}\,\,\, 10$, $\Delta E(t)$ 
is positive corresponding to LTP described by the
experimental data~\cite{rf:2}.
For $E(t) \,\,\,\vphantom{(}\lower4pt\hbox{$\stackrel
{ \displaystyle {>}}{\sim}$}\,\,\, 20$ and $S(t) \,\,\,\vphantom{(}\lower4pt\hbox{$\stackrel
{ \displaystyle {>}}{\sim}$}\,\,\, 10$, $\Delta E(t)$ is zero
due to the saturation effect.  For $E(t) \,\,\,\vphantom{(}\lower4pt\hbox{$\stackrel
{ \displaystyle {<}}{\sim}$}\,\,\, 15$ and $S(t) \,\,\,\vphantom{(}\lower4pt\hbox{$\stackrel
{ \displaystyle {<}}{\sim}$}\,\,\,
10$, $\Delta E(t)$ is zero because of less $Ca^{2+}$
entry.  For $E(t) \,\,\,\vphantom{(}\lower4pt\hbox{$\stackrel
{ \displaystyle {>}}{\sim}$}\,\,\, 15$ and $S(t) \,\,\,\vphantom{(}\lower4pt\hbox{$\stackrel
{ \displaystyle {<}}{\sim}$}\,\,\, 10$, $\Delta E(t)$
is negative corresponding to LTD 
at low frequency~\cite{rf:6}.

When we consider $N$ neurons which are stimulated independently, 
the averaged output $<O(t)>$ at time $t$ is defined by
\begin{equation}
<O(t)> = \frac{1}{N}\sum_{i=1}^{N}O_{i}(t),
\end{equation}
which can be written as 
\begin{equation}
<O(t)> \sim f(V(t))
\end{equation}
in a large $N$ limit according to the central limit theorem.
Therefore, in this letter we consider
$f(V(t))$ to represent
the characteristic measure of the LTP amplitude.
In Figs. 2 - 4, the amplitude
variable of the vertical axis shows $<<f(V(t))>_{t}>_{e}$ for
comparison with the physiological data in ref. 2; the notation
$<\cdot>_{t}$ represents the time average for a sample process and
$<\cdot>_{e}$ represents the ensemble average for a large number of
sample processes.  In the present letter, the number of
ensembles is fixed at $10^3$ to $10^5$.  The
values of the parameters used in this letter are chosen to be
biologically plausible, i.e., $V_{th}=-55$ mV, $V_{r} = -70$ mV, 
$\delta(t)$ is
generated from the uniform distribution of $-E(0)/100 \leq
\delta(t) \leq E(0)/100$, $E(0) = 15$ mV, $C = 0.15$, $V_{L} =
-60$ mV, $\alpha = 10$ ms, $\beta = 200$ ms, $C_{1} = 0.00015,
C_{2} = 0.01, C_{3} = 8, C_{4} = 0.25, C_{5} = 20, C_{6} =
4, C_{7} = 0.5, C_{8} = 0.25, C_{9} = 5$, and $C_{10} = 1$.

\section{Time Structure of the Input Stimuli}
To determine the effect of the spike structure on LTP, we consider three
different spike stimuli: (i) regular, (ii) Markov, and (iii)
chaotic.

For the Markov stimuli, we use the same transition matrix as was 
used by Tsukada
et al.~\cite{rf:2} to compare our results with theirs; 
it is given as
\begin{equation}
P = [P_{ij}] = \left[ \begin{array}{ccc}
SS & SM & SL \\
MS & MM & ML \\
LS & LM & LL
\end{array} \right],
\end{equation}
where $S$, $M$, and $L$ stand for the Markovian states of
the signals with a 100 ms spike interval, 500 ms spike
interval, and 900 ms spike interval, respectively.  For 
example, the transition probability $P_{ij} = MS$ indicates
that the $S$ (100 ms) interval directly follows the $M$ 
(500 ms)
interval.
The transition matrices used in the simulations are given as
\begin{equation}
P_{p} = [P_{ij}] = \left[ \begin{array}{ccc}
0.70 & 0.20 & 0.10 \\
0.45 & 0.10 & 0.45 \\
0.10 & 0.20 & 0.70
\end{array} \right]
\end{equation}
\begin{equation}
P_{i} = [P_{ij}] = \left[ \begin{array}{ccc}
0.41 & 0.18 & 0.41 \\
0.41 & 0.18 & 0.41 \\
0.41 & 0.18 & 0.41
\end{array} \right]
\end{equation}
\begin{equation}
\hspace{5pt}P_{n} = [P_{ij}] = \left[ \begin{array}{ccc}
0.10 & 0.20 & 0.70 \\
0.45 & 0.10 & 0.45 \\
0.70 & 0.20 & 0.10
\end{array} \right],
\end{equation}
where $P_{p}$, $P_{i}$, and $P_{n}$ correspond to the positive correlation, the
independent correlation, and the negative correlation,
respectively~\cite{rf:2}.  The mean frequency of the input
stimuli $\sigma (t)$ for all
three cases is fixed at 2 Hz.

For the case of chaotic stimuli, we use the modified Bernoulli
map $x_{n} \rightarrow x_{n+1}$ (mod.1) to produce the input
stimuli $\sigma (t)$,
\begin{equation}
x_{n+1} = \left \{ \begin{array}{ll}
x_{n}+2^{B-1}(1-2\epsilon)x_{n}^{B}+\epsilon, & (0
\leq x_{n} \leq 1/2) \\
x_{n}-2^{B-1}(1-2\epsilon)(1-x_{n})^{B}-\epsilon, & (1/2 < x_{n}
\leq 1),
\end{array} \right.
\end{equation}
where $ 1 \leq B \leq 3$ and $\epsilon = 10^{-13}$ are
parameters.  Then, in order to fix the mean spike interval at
500 ms (i.e., the mean frequency is 2 Hz),
we transform the value of $x_{n}$ to the spike interval
$\tau$ of the time
course $\sigma (t)$ in the following way:
\begin{equation}
\left \{ \begin{array}{ccc}
0 \leq x_{n} < 1/9 & \rightarrow & \tau = 100ms \\
1/9 \leq x_{n} < 2/9 & \rightarrow & \tau = 200ms \\
2/9 \leq x_{n} < 3/9 & \rightarrow & \tau = 300ms \\
3/9 \leq x_{n} < 4/9 & \rightarrow & \tau = 400ms \\
4/9 \leq x_{n} < 5/9 & \rightarrow & \tau = 500ms \\
5/9 \leq x_{n} < 6/9 & \rightarrow & \tau = 600ms \\
6/9 \leq x_{n} < 7/9 & \rightarrow & \tau = 700ms \\
7/9 \leq x_{n} < 8/9 & \rightarrow & \tau = 800ms \\
8/9 \leq x_{n} \leq 1 & \rightarrow & \tau = 900ms.
\end{array} \right.
\end{equation}
The properties of this
map were thoroughly studied by Aizawa,~\cite{rf:7} and it should be
noted that the map is stationary chaotic for $B < 2$, and
nonstationary chaotic for $B \geq 2$.  The time
course of the input stimuli $\sigma (t)$ generated by eq. (3.6)
reveals a typical
intermittency for all regimes, and the $1/f$ 
spectrum is emphasized in the case for $B \geq 2$.

\section{Numerical Simulations}
Let us now numerically investigate LTP induced by the
time-structured stimuli mentioned above.
The setup of the numerical LTP experiments is as follows.
The unit of the discrete time is taken as 1 ms, and 20 impulses with
0.05 Hz are applied as an initial control stimulus.  Then, we
ensure that the
amplitude $E(t)$ reveals a controlled reference state corresponding
to the threshold $V_{th}$.  After these preparations, 200
tetanic impulses are applied, and then 20 control impulses with
0.05 Hz intervals (called the latter control) are applied.  
The LTP amplitude is measured in terms of the average value,
$<f(V(t))>_{t}$, where $t$ stands for the duration of the
latter control.  These procedures for our
simulations
are the same as in the physiological experiments~\cite{rf:2}.

The first case is that of the regular periodic stimuli ranging
from 10 Hz to 0.5 Hz, and the results are shown in Fig. 2.  We use 1,000
different sample processes of 
stimuli for each frequency, and the ensemble averages of LTP,
$<<f(V(t))>_{t}>_{e}$, are shown 
by the filledcircles and the standard
deviations by the errorbars.
Figure 2 clearly shows that a higher-frequency stimulus produces
a larger LTP than lower-frequency ones, 
and that no LTP occurs at frequencies of 
less than 2 Hz.  These results are consistent with the
results of experiments performed 
by Tsukada et al.~\cite{rf:2}

\begin{figure}
\epsfxsize=15.9cm \epsfysize=10.0cm \epsfbox{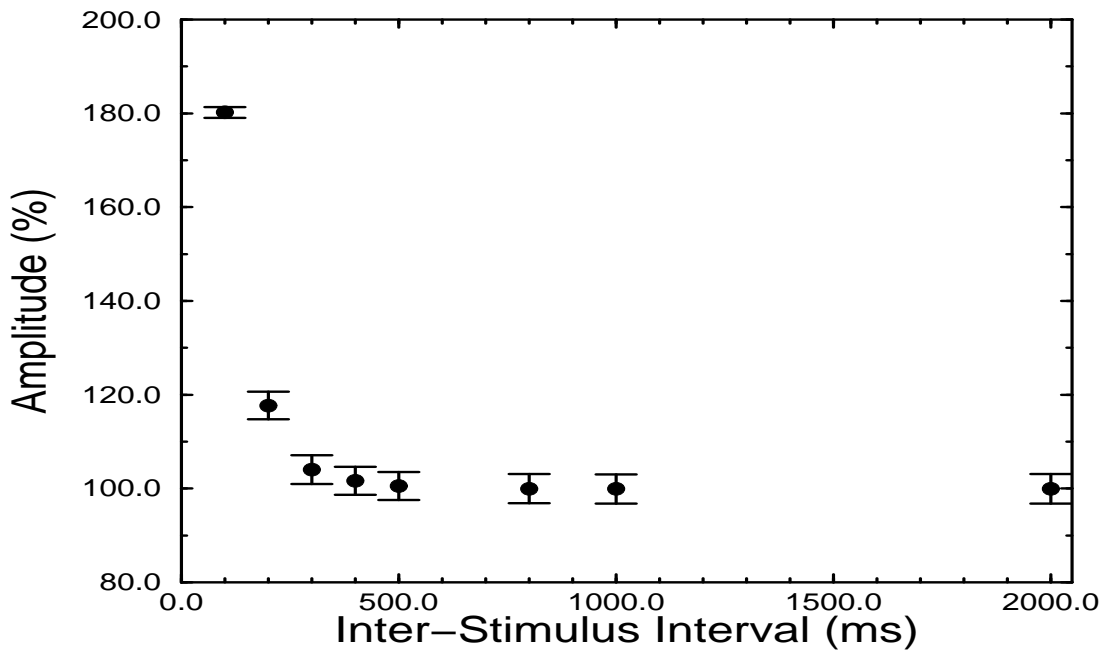}
\caption{Amplitude of LTP induced by regular stimuli.
One thousand initial 
conditions are applied to each frequency, and the
filledcircles and the errorbars represent the averages and the
standard deviations, respectively.}
\end{figure}

Next, we apply the Markov stimuli produced by the transition
matrices in
eqs. (3.1), (3.2), and (3.3).  For each type of
stimulus, we consider 100,000 sample processes, and
calculate the average and standard
deviation of the LTP amplitude.
The results are shown in Fig. 3.  It is clear that the positive
correlated stimuli produce the largest LTP, the independent ones
the next largest ones, and the negative correlated ones the
least ones.  Comparing
Fig. 3 and Fig. 2, we can conclude that the time-structured Markov
stimuli with 2 Hz frequency produce a larger
LTP than do the regular stimuli with the same frequency.
This implies that the time structure of the input signals is 
important for LTP to occur.
Although our result for the independent correlation case
does not coincide with the experimental result obtained by Tsukada
et al.,~\cite{rf:2}
the positive and the negative cases fit well with their
experimental results.
Furthermore, our results shown in Figs. 2 and 3 are consistent
with the results of previous work performed by Sugiura et
al.~\cite{rf:8} with a different neuron model.

\begin{figure}
\epsfxsize=15.9cm \epsfysize=10.0cm \epsfbox{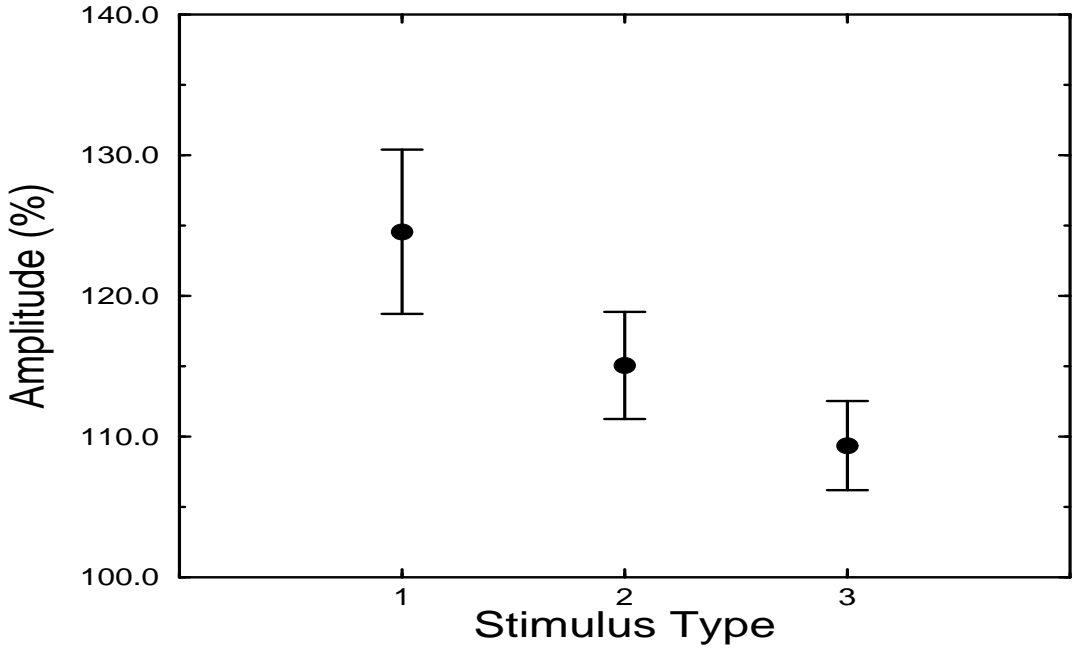}
\caption{Amplitude of LTP induced by Markov stimuli.  The stimulus
types 1, 2, and 3 correspond to the positive, the independent,
and the negative correlation, respectively.  One hundred
thousand initial
conditions are applied to each type, and the filledcircles
and the errorbars represent the averages and the standard deviations, 
respectively.}
\end{figure}

Finally, we apply the chaotic intermittent stimuli expressed 
by eq. (3.6) to our
neuron model.  For each value of $B$, we produce 10,000
sample processes.  The results are shown in
Fig. 4.  As the value of $B$ increases, the average LTP
amplitude increases almost
monotonically.  Thus we can conclude that on average, the
chaotic stimuli in the nonstationary regime produce larger
LTP than do the stationary
ones.  Comparing Fig. 4 with Figs. 2 and 3, we can clearly
see that intermittent chaotic stimuli have the strongest LTP 
effect among all the stimuli, though the fluctuations
represented by the standard deviations are also much larger
in the nonstationary regime than in the stationary regime.

\begin{figure}
\epsfxsize=15.9cm \epsfysize=10.0cm \epsfbox{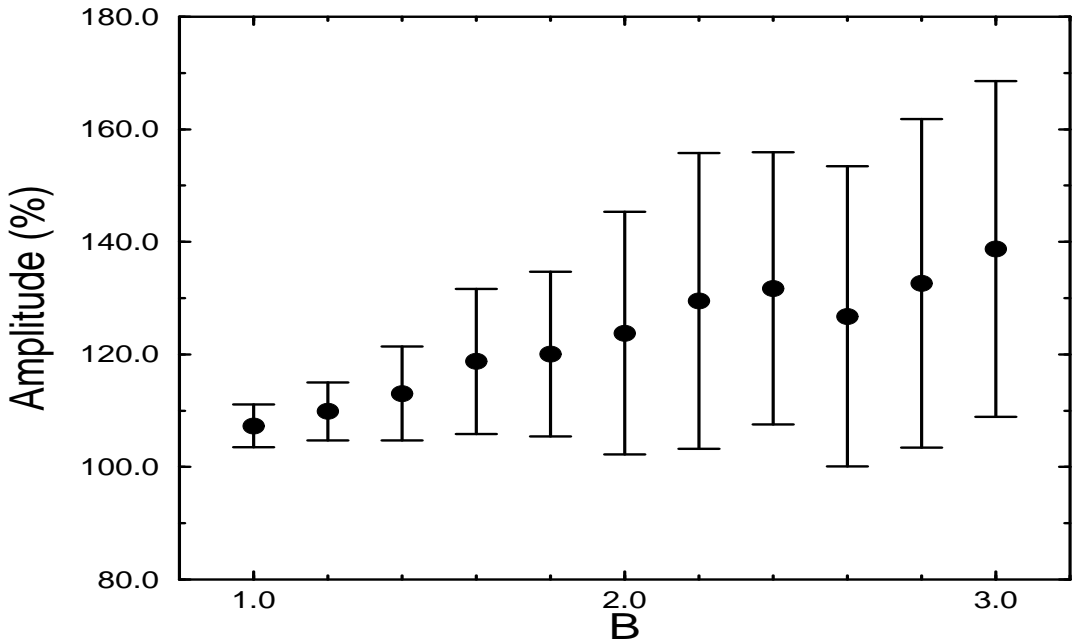}
\caption{Amplitude of LTP induced by intermittent chaotic
stimuli.  Ten thousand
initial conditions are applied to each B value, and the
filledcircles and the errorbars represent the averages and
the standard 
deviations, respectively.}
\end{figure}

\section{Discussion}
In this letter, we proposed a simple neuron model
including the effects of NMDA and non-NMDA
receptors, and discuss the relation between the
time structure of the input stimuli and the amplitude of
LTP.  It is shown
by numerical
simulation that the results obtained using our neuron model
fit the experimental data reasonably well, and indicate
that the chaotic stimuli in the nonstationary regime will
produce the largest LTP.

The reason that the time structure of the input stimuli has
a significant effect on the amplitude of LTP can be understood
by the fact that the calcium concentration through NMDA
receptors has a long-lasting memory effect ($\beta=200$ ms),
as is shown in eq. (2.5).  Thus, if the interval between two
spikes is short, the effect of the first spike remains when 
the second one arrives, resulting in a higher calcium
concentration in a post-synaptic neuron.  On the other hand,
when the interval between the two spikes is long, the effect of
the first spike ceases before the second one arrives.
Therefore, the inter-spike-interval must be short
for LTP to occur in our model.  For the
Markov stimuli, the positive correlation case has the
longest short spike interval, the
independent one the next longest, and the negative one the
shortest,
and this explains the differences in the average LTP
amplitude in Fig. 3.
In the case for the chaotic
stimuli, $x_{n}$ of the modified Bernoulli map tend
to localize around $x_{n} \sim
0$ or $x_{n} \sim 1$ in the large $B$ regime, and as the
value of $B$ increases, the spikes
tend to cluster in a short interval or a long interval for a
longer duration.  Therefore, if the spikes occur within a short
interval for a long duration, they will produce very large
LTP.  On the other hand, almost no LTP will occur when
spikes occur within a
long interval for a long duration.  This results in the large
deviation in the nonstationary regime in Fig. 4.
Since the short spike interval becomes longer as the value
of $B$ increases, the
average LTP amplitude also increases.
Moreover, the average LTP amplitude in the
nonstationary chaotic regime is the largest among those for all
the stimuli, because 
the short spike interval in the
nonstationary chaotic regime is longer than that in any
others.

To determine how memory information is represented in the
brain, it is
necessary to elucidate the relationship between the time structure of
the impulse train and its validity as a form of information representation.
For this purpose, we obtained the basic data of the LTP formation 
at the single neuron level.  Our results
strongly suggest that the time structure of the impulse train is
important for memory information representation at the single
neuron level.  In particular, the result that the
nonstationary chaotic spike trains produce the largest LTP 
implies that the brain may use
chaotic intermittency for memory
information representation even at the single neuron level.
Furthermore, the wide range of the LTP amplitude in the
nonstationary regime, which is indicated by the large
standard deviation, seems to play a very important role in
the selection of the detailed information embedded in the
diversity of the time-structured spike trains.

However, in order to acquire the entire pictures of memory information
representation and its algorithm in the brain, it is also
necessary to determine
the relationship between the time structure of the impulse train and
its validity as a form of information representation at the network
level.  Also, in particular, detailed
study about the relationship between
the time structure of the impulse train and the properties of the
network connections,
such as symmetric, asymmetric, excitatory, and inhibitory ones, is
necessary.  To this end, the neuron model proposed in
this letter is suitable because it is simple enough for
network-level simulation and has biological
properties, though some details such as the form of
eq. (2.6) can be further simplified.  Further research
for analysis of the network structure and increasing the
level of the sophistication of our neuron model is now in progress.

\end{document}